\documentclass[11pt]{revtex4}

\usepackage{graphicx}

\begin{document}
\title{ Quantum Phases of Ultracold Bosonic Atoms in a Two-Dimensional Optical Superlattice }
\author{Jing-Min Hou\footnote{Electronic address:
jmhou@seu.edu.cn}} \affiliation{Department of Physics, Southeast
University, Nanjing, 211189, China}

\begin{abstract}
\centerline{\bf Abstract}

 We study quantum phases of ultracold bosonic atoms in a
two-dimensional optical superlattice. The extended Bose-Hubbard
model derived from the system of ultracold bosonic atoms in an
optical superlattice is solved numerically with  Gutzwiller
approach. We find that  the modulated superfluid(MS),
Mott-insulator (MI) and density-wave(DW) phases appear in some
regimes of parameters. The experimental detection of the first
order correlations and the second order correlations of different
quantum phases with time-of-flight and noise-correlation
techniques is proposed.

\noindent{\it PACS}: 03.75.Lm, 03.75.Nt, 32.80.Pj\\
{\it keywords}: Optical superlattice; Ultracold atom; Quantum
phase

\end{abstract}
\maketitle

 The system of ultracold atoms in optical
lattices, which provides an intriguing environment to study
strongly correlated condensed matter systems and quantum
information due to  the complete control over the system
parameters, has been extensively investigated theoretically  and
experimentally. The superfluid-Mott-insulator transition of
ultracold bosons in optical lattices has been theoretically
analyzed and experimentally demonstrated\cite{Jaksch,Greiner}. The
realization of 1D quantum liquids with atomic gases has
achieved\cite{Paredes, Stoferle}. A Bose glass phase has been
observed in  an optical lattice\cite{Fallani}. Repulsively bound
states in an optical lattice have been observed in a recent
experiment\cite{Winkler}. Atomic Bose-Fermi mixtures in an optical
lattice have been studied\cite{Gunter,Ospelkaus}. Ultracold
fermionic atoms in optical lattices have also been investigated
recently\cite{Zwierlein, Chin,Kohl,Stoferle2,Rom}. Magnetic
quantum phase transition,  nonlinear dynamics of a dipolar
Bose-Einstein condensate, phase diagram of two-species
Bose-Einstein condensates and magnetic soliton and soliton
collisions of spinor Bose-Einstein condensates in an optical
lattice have been studied\cite{He,Xie,Li,Zheng}. Diverse schemes
of
 quantum information processing in optical lattices have been
 proposed\cite{Jaksch1,Brennen,Pachos}.
The quantum phases of ultracold bosons in one-dimensional optical
superlattices have been discussed in some
literatures\cite{Buonsante,Buonsante2,Buonsante3,Buonsante4,Buonsante5,Rousseau}.
However, the system of ultracold bosons in  a  two-dimensional
optical superlattice has not investigated as yet.

In this paper, we will study the system of ultracold bosonic atoms
in a two-dimensional optical superlattice. We will show that novel
quantum phases, such as modulated superfluid, density wave, and
Mott-insulator phases, appear  in some regimes of quantum
parameters. We will display the first order correlation function
and the second order correlation function in the lattice momentum
space for different quantum phases in order to identify them in
experiments with the time-of-flight and noise-correlation
techniques.

\begin{figure}[ht]
 \includegraphics[width=0.99\columnwidth]{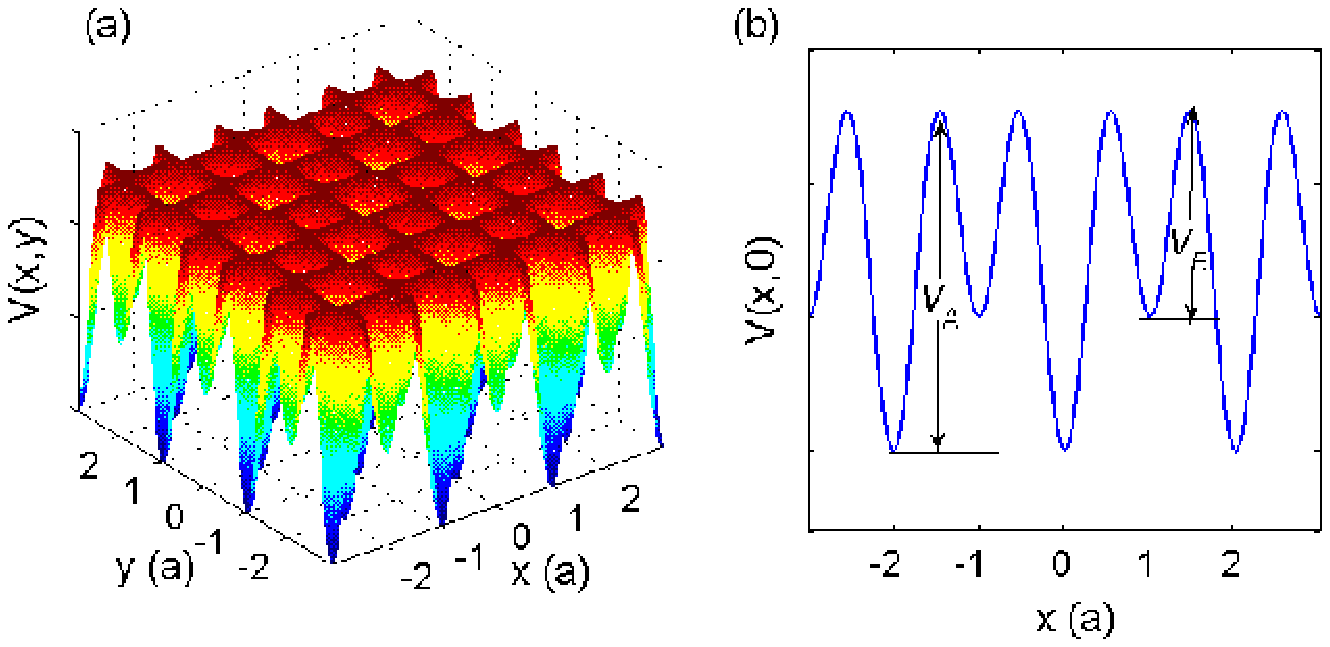}
\caption{(color online). (a) Schematic diagram of a
two-dimensional optical superlattice potential, where $x$ and $y$
are coordinates
 in units of lattice
 costant $a$ and $V(x,y)$ represents the potential energy  at position $(x,y)$.
  (b) The profile of the optical superlattice potential shown in (a) at $y=0$. Here $V_A$ and $V_B$
 represent the lattice depths of the sublattices A and B respectively. } \label{fig1}
\end{figure}

 Ultracold bosons with repulsive interactions in an
 external potential can be described by a second quantized Hamiltonian
 as follows,
\begin{eqnarray}
H&=&\int d^3 r\psi^\dag({\bf r})\left[-\frac{\hbar^2}{2m}{\bf
\nabla}^2+V({\bf r}) -\bar\mu\right]\psi({\bf r}) +\frac{g}{2}\int
d^3 r\psi^\dag({\bf r})\psi^\dag({\bf r})\psi({\bf r})\psi({\bf
r}) \label{hamiltonian1}
\end{eqnarray}
where $\psi({\bf r})$ and $\psi^\dag({\bf r})$ are the boson field
operators for annihilating and creating an atom at site ${\bf r}$
respectively; $g=4\pi a_s \hbar^2/m$, where $a_s$ is the $s$-wave
scattering length, which is positive for atoms with repulsive
interactions; $m$ is the mass of  atoms and $\bar{\mu}$ is the
chemical potential;  $V({\bf r})$ is an external potential. In our
model, the external potential   $V({\bf r})$  is a two-dimensional
superlattice potential such as
$V(x,y)=V_1\cos^2(kx)\cos^2(ky)+V_2[\cos^2(kx/2)\cos^2(ky/2)+\sin^2(kx/2)\sin^2(ky/2)]$,
which is schematically displayed in FIG.\ref{fig1}. This
superlattice potential consists of two sublattices that we denote
as sublattice `A' and sublattice `B' with the potential depths
$V_A$ and $V_B$, respectively. Such a two-dimensional superlattice
can be built with optical techniques. For simplicity, we do not
consider the shallow harmonic trapping potential, which is usually
required in experiments to trap atomic gases.

In periodic potentials, the atomic states are represented by Bloch
wave functions, which can be expanded in terms of a set of Wannier
states that are well  localized at lattice sites. Because of the
translational symmetry of each sublattice, the Wannier function
$w_i({\bf r})$ is equal to $w_A({\bf r}-{\bf R}_i)$  for
sublattice A and is equal to $w_B({\bf r}-{\bf R}_i)$  for
sublattice  B, where ${\bf R}_i$ is a lattice vector.
 At zero temperature, the atoms are in the lowest band states. Expanding
the field operators in the lowest vibrational Wannier states as
$\psi({\bf r})=\sum_{i} b_{i}w_{i}({\bf r})$,
Eq.(\ref{hamiltonian1}) reduces to the extended Bose-Hubbard
Hamiltonian,
\begin{eqnarray}
H&=&-t\sum_{i}\left[b_{(i_x,i_y)}^\dag
b_{(i_x+1,i_y)}+b_{(i_x,i_y)}^\dag b_{(i_x,i_y+1)}+{\rm
H.C.}\right]-\sum_{i}\mu_{i} n_{i} +\frac{1}{2}\sum_{i}U_{i}
n_{i}(n_{i}-1)
 \label{hamiltonian2}
\end{eqnarray} where
$(i_x,i_y)$ is the coordinate  position of lattice site $i$ in the
two-dimensional optical superlattice.  $b_{(i_x,i_y)}$ and
$b^\dag_{(i_x,i_y)}$ represent the operators for annihilating and
creating an atom at lattice site $i=(i_x,i_y)$ respectively;
$t=\int d{\bf  r}\ w^*_A({\bf r}-{\bf
r}_i)[-\frac{\hbar^2}{2m}\nabla^2+V({\bf r})]w_B({\bf r}-{\bf
r}_j)$ is the tunnelling parameter between
 adjacent lattice
sites in the different sublattices A and B. $U_{i}$ is the
repulsive on-site interaction between atoms at lattice site $i$;
$\mu_{i}$ is the effective chemical potential corresponding to
lattice site $i$. In the two different sublattices, the effective
chemical potential and the on-site interaction are
\begin{eqnarray*}
\cases{\mu_{i}=\mu, U_{i}=U_A& when $i_x+i_y$ is even,\cr
\mu_{i}=\mu-\delta, U_{i}=U_B&when $i_x+i_y$ is odd,}
\end{eqnarray*}
where $\mu=\bar{\mu}-\int d{\bf r}V({\bf r})|w_A({\bf r})|^2$ and
$\delta=\int d{\bf r}V({\bf r})(|w_B({\bf r})|^2-|w_A({\bf
r})|^2)$; $U_A=g\int d{\bf r}|w_A({\bf r})|^4/m$ and $ U_B=g\int
d{\bf r}|w_B({\bf r})|^4/m$ are the on-site interactions on the
two different kinds of sublattice sites.

 With Gutzwiller's ansatz, we can
write the wavefunctions of the system as\cite{Jaksch},
\begin{eqnarray}
|\Psi_G\rangle=\prod_i|\phi_i\rangle,\ \ \ \
|\phi_i\rangle=\sum_nf^{(i)}_n|n\rangle_i,
\end{eqnarray}
where $|n\rangle_i$ is the Fock state with $n$ particles at the
lattice site $i$ and $f_n^{(i)}$ are the variational parameters,
which satisfy the normalized condition $\sum_n |f_n^{(i)}|^2=1$
for each lattice site $i$. To find the ground state of the system,
we minimize the expectation value of the
Hamiltonian(\ref{hamiltonian2})$, \langle\Psi_G|H|\Psi_G\rangle$,
with the variational parameters $\{f_n^{(i)}\}$, under a given
 chemical potential $\mu$. In our work, we use conjugate
gradient algorithm to minimize the expectation value of the
Hamiltonian. The system contains $10\times 10$ lattice sites and
the truncated number of Fock states is $6$. To eliminate the
boundary effects,  periodic boundary conditions are used.

 To identify different quantum phases,
we define the static structure factor as follows:
\begin{eqnarray}
S({\bf k})=\frac{1}{N}\sum_{i,j}e^{i{\bf k}\cdot({\bf R}_i-{\bf
R}_j)}\langle n_in_j\rangle
\end{eqnarray}
where ${\bf R}_i$ is a lattice vector and $N$ is the number of
lattice sites. The diagonal long-range order  and the off-diagonal
long-range order in the system of ultracold bosonic atoms in a
two-dimensional optical superlattice are measured by $S(\pi, \pi)$
and $\phi=\langle b\rangle$, respectively. The Mott-insulator
phase is a state with particle number density pinned at an integer
such as $n=1,2,\cdots$, which corresponds to a commensurate
filling of the lattice. The Mott-insulator phase is characterized
by $S(\pi, \pi)= 0$ and $\phi= 0$.  In the density wave phase, the
particle number density is modulated by with the period of twice
lattice constant  and is pinned at two different integers in the
two different sublattices. The density wave phase is a
superlattice crystal with the diagonal long-range order measured
by $S(\pi, \pi)$, and without the off-diagonal long-range order.
Therefore, $S(\pi, \pi)\neq 0$ and $\phi= 0$ feature the density
wave phase. In the modulated superfluid phase, the diagonal
long-rang order and the off-diagonal long-range order coexist, so
the modulated superfluid phase is characterized by $S(\pi,
\pi)\neq 0$ and $\phi\neq 0$. The conventional superfluid phase,
which is absent in our model, is characterized by $S(\pi, \pi)= 0$
and $\phi\neq 0$.

\begin{figure}[ht]
 \includegraphics[width=\columnwidth]{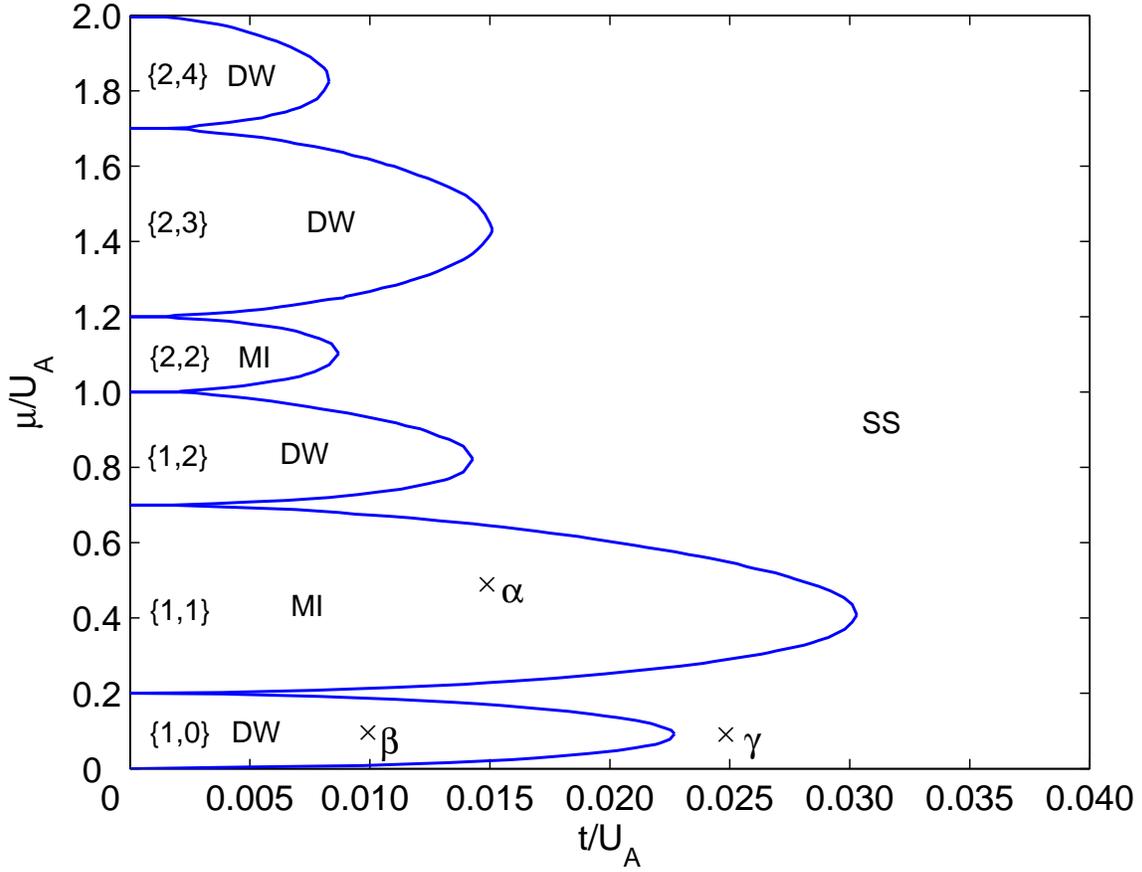}
\caption{Phase diagram of ultracold bosonic atoms in an optical
superlattice for $U_B=0.5 U_A$ and $\delta=0.2U_A$. $t/U_A$ and
$\mu/U_A$
  are the dimensionless hopping matrix element and the chemical potential respectively.  In this
  diagram, MI, DW and MS denote the Mott-insulator, density wave and modulated superfluid phases respectively.
  $\{n_A, n_B\}$ represents the average particle density in sublattices A and B when
  the system is in MI or DW phase. The points marked with `$\times$', of which the first order
   correlation function and the second
  order correlation function are shown in FIG.\ref{fig3}, are denoted by `$\alpha$', `$\beta$',
  and `$\gamma$'.}\label{fig2}
\end{figure}

FIG. \ref{fig2} shows the phase diagram of ultracold bosonic atoms
in an optical superlattice for $U_B=0.5 U_A$ and $\delta=0.2U_A$,
which is obtained by solving the Hamiltonian (\ref{hamiltonian2})
numerically with Gutzwiller approach. This phase diagram is a
result of competition among the tunnelling term and the two
different on-site interactions in the two different sublattices.
From this figure, we can find that Mott-insulator, density wave
and modulated superfluid phases appear in some regimes of
parameters. The Mott-insulator and density wave phases are denoted
as $\{n_A, n_B\}$ for their filling numbers in the sublattice A
and B. The second and fourth lobes counted from the bottom,
denoted as $\{1,1\}$ and $\{2,2\}$ respectively,  belong to the
Mott-insulator phase, For the Mott-insulator phase $\{1,1\}$, the
critical value of $t/U_A$ for the MI-MS phase transition is 0.031,
while the corresponding value of the MI-SF transition in a
two-dimensional optical lattice is 0.043\cite{Jaksch}. Thus, the
Mott-insulator phase is more instable in optical superlattice than
in optical lattice due to competition between the different
on-site interactions in the different sublattice in the optical
superlattice. The first, third, fifth  and sixth lobes, denoted as
$\{1,0\}$, $\{1,2\}$, $\{2,3\}$ and $\{2,4\}$ respectively, belong
to the density wave phase. The reason that the density wave phase
$\{1,0\}$ appears is the difference of the effective chemical
potentials for the two different sublattice, while the other
density wave lobes are the result of competition among the
tunnelling term and the two on-site interaction in the two
sublattices. The other regime except the Mott-insulator and
density wave lobes, where the diagonal long-range order and the
off-diagonal long-range order appear simultaneously,  belongs to
the modulated superfluid phase. In our model, the superfluid
phase, which conventionally appears in the optical lattice-cold
atom system\cite{Jaksch,Greiner}, is modulated into the modulated
superfluid phase by the periodic modulation of the optical
superlattice.

 The techniques to detect the above
quantum phases are accessible for now. Altman \em et al\em. have
proposed to utilize shot-noise correlations to probe complex
many-body states of trapped ultracold atoms, e.g., Mott
states\cite{Altman}. Scarola \em et al\em.   have suggested to use
the noise-correlation technique to detect the supersolid phase of
cold atoms with long-range dipolar interactions in an optical
lattice\cite{Scarola}. The correlations for the case of a bosonic
Mott insulating state\cite{Rom},  and the pair-correlated atoms
created by dissociating weakly bound diatomic molecules near a
Feshbach  resonance\cite{Greiner1} have recently been observed
through shot-noise correlations techniques experimentally.

For the time of flight detection, the atoms interact weakly after
the trapped optical lattice is
 adiabatically turned off, and the number
 of atoms at position ${\bf r}$  in the expanding cloud after a time $T$
 is given by $\langle n({\bf r})\rangle_T\approx|\tilde{w}({\bf k}({\bf r}))|^2\rho_0({\bf k}({\bf
 r}))$, where $\tilde{w}$ is the Fourier transform of the Wannier
 function\cite{Altman,Scarola}.
 The relation between  the lattice momentum  and the position in the expanding
 cloud after an expanding time $T$
 is ${\bf k}({\bf r})=a_0m{\bf r}/\hbar T$, where $m$ is the mass
 of  atom and $a_0$ is the width of the Wannier state
in the lattice. $\rho_0$ is a first order correlation function in
the lattice momentum space which is defined as follows,
\begin{eqnarray}
 {\rho}_0({\bf k})=\sum_{i,j} e^{i{\bf k}\cdot({\bf R}_i-{\bf
R}_j)}\langle b_i^\dag b_j\rangle,
\end{eqnarray}
where ${\bf R}_i$ is a lattice vector. The first order normalized
correlation function is defined as $\rho({\bf k})\equiv\rho_0({\bf
k})/N$, where $N$ is the number of lattice sites. The shot-noise
correlation function in the expanding atomic cloud after an
expanding time $T$ is ${\cal G}({\bf r},{\bf r}')=\langle n({\bf
r})n({\bf r}')\rangle_T-\langle n({\bf r})\rangle_T\langle n({\bf
r}')\rangle_T$, which is proportional to momentum correlations in
the ground state of the optical lattice trapped system,
\begin{eqnarray}
{\cal G}({\bf r},{\bf r}')\sim |\tilde{w}({\bf k}({\bf
r}))|^2|\tilde{w}({\bf k}({\bf r}'))|^2G_0({\bf k}({\bf r}),{\bf
k}({\bf r}')),
\end{eqnarray}
where $G_0({\bf k},{\bf k}')$ is  the second order correlation
function in the lattice momentum space. the second order
normalized correlation function is defined by $G({\bf k},{\bf
k}')=G_0({\bf k},{\bf k}')/\rho_0({\bf k})\rho_0({\bf k}')$, which
can be written as \cite{Scarola},
\begin{eqnarray}
G({\bf k},{\bf k}')&=&\frac{\sum_{ii'jj'} e^{i{\bf k}\cdot ({\bf
R}_i-{\bf R}_{i'})+i{\bf k}'\cdot( {\bf R}_j-{\bf R}_{j'})}\langle
b_i^\dag b_j^\dag b_{j'}b_{i'}\rangle}{\rho_0({\bf k})\rho_0({\bf
k}')}-1+\frac{ma_0}{\hbar T}\frac{\delta({\bf k}-{\bf
k}')}{|\tilde{w}({\bf k}')|^2\rho_0({\bf k}')},
\end{eqnarray}
where the $\delta({\bf k}-{\bf k}')$ term is the autocorrelation
term and will be dropped in the subsequent discussion because it
only contributes to the second order correlation function  for
${\bf k}={\bf k}'$ and does not contain the characterizing
information about the correlation function distribution in the
momentum space.

\begin{figure}[ht]
 \includegraphics[width=0.9\columnwidth]{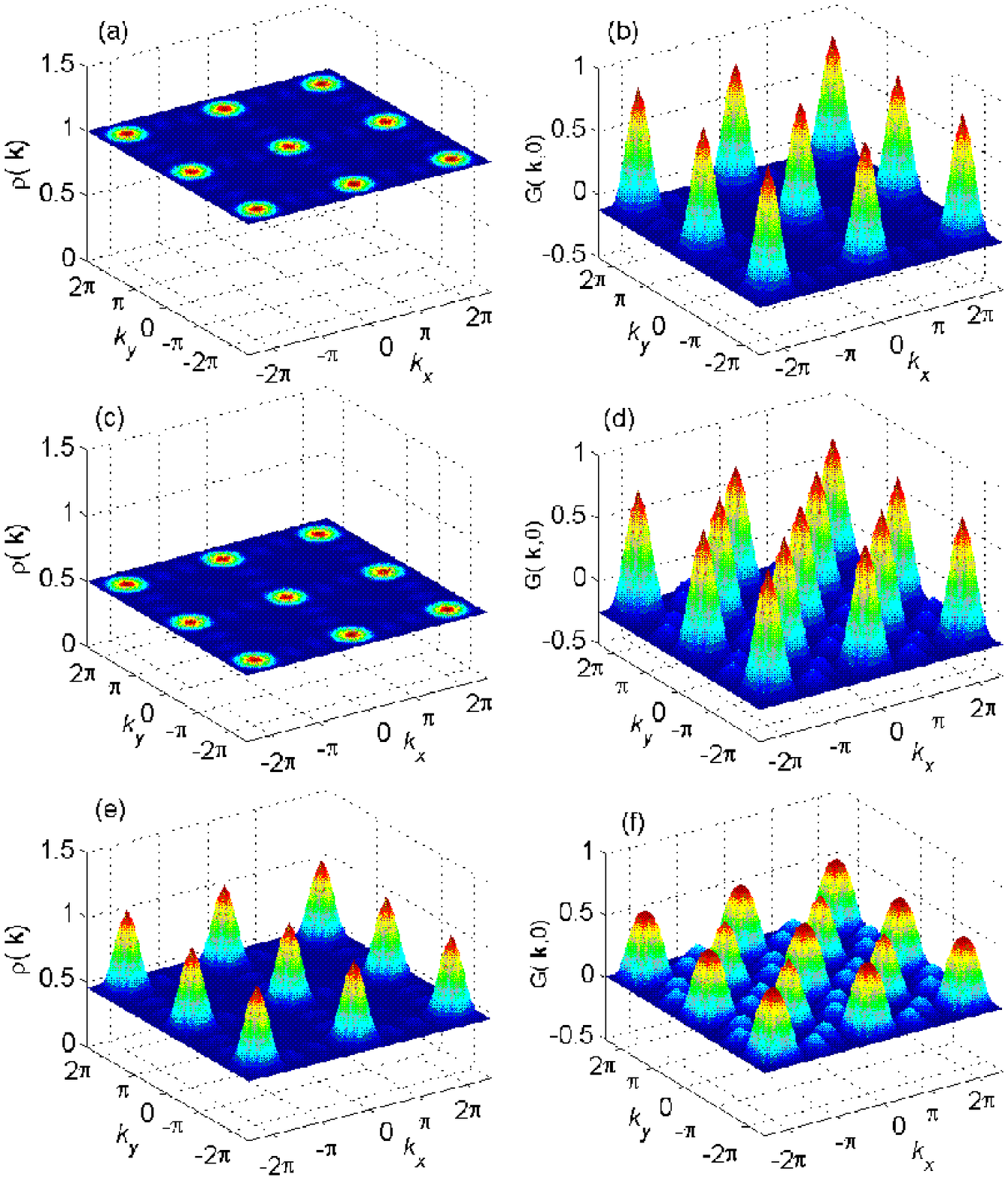}
\caption{(color online). The first order and the second order
normalized density correlations of different quantum phases
  as functions in the lattice momentum space. (a) and (b) are respectively the first order and the second order
normalized density correlations in the lattice momentum space of a
Mott-insulator phase at the point denoted by `$\alpha$'  in
FIG.\ref{fig2}. (c) and (d) are respectively the first order and
the second order normalized density correlations in the lattice
momentum space of a density wave phase at the point denoted by
`$\beta$'  in FIG.\ref{fig2}. (e) and (f) are respectively the
first order and the second order normalized density correlations
in the lattice momentum space of a modulated superfluid phase at
the point denoted by `$\gamma$'  in FIG.\ref{fig2}. } \label{fig3}
\end{figure}

To identify the quantum phases, such as the Mott-insulator,
density wave and modulated superfluid phases, in experiments, we
pick out three points $\alpha, \beta$ and $\gamma$, representing
the three quantum phases respectively, from the phase diagram
shown in FIG.\ref{fig2}. For the three picked  points, we
  calculate
the distributions of  $\rho({\bf k})$ and  $G({\bf k},0)$ in the
lattice momentum space,
 which are shown in FIG. \ref{fig3}.
 The first order normalized correlation function is almost a
constant  for the Mott-insulator  and density wave phases, while,
for the modulated superfluid phase, it has peaks at points
$(0,0)$, $(0,\pm 2\pi)$, $(\pm 2\pi,0)$ and $(\pm 2\pi,\pm 2\pi)$
in the lattice momentum space, which indicate that the
off-diagonal long-range order exists. The second order normalized
correlation function has peaks at points $(0,0)$, $(0,\pm 2\pi)$,
$(\pm 2\pi,0)$ and $(\pm 2\pi,\pm 2\pi)$ in the lattice momentum
space for the Mott-insulator, while, for the density wave and
modulated superfluid phases, it has additional peaks at $(\pm
\pi,\pm \pi)$, which indicate that the diagonal long-range order
exists. These characterizing features of correlation functions in
the lattice momentum space can be detected by probing the
correlations in the coordinate space of the atomic cloud after an
expanding time $T$ with the time-of-flight and noise-correlation
techniques.

In our work, for simplicity, we do not consider the shallow
harmonic trapping potential, which is usually  required to trap
atomic gases in the practical experiments. In fact, the presence
of a shallow harmonic trapping potential does not change the
intrinsic features that we have discussed. The energy offset due
to the shallow harmonic trapping potential can be combined in the
effective chemical potential. Thus,  the effective chemical
potential varies with the position, so the phase separation may
occurs in the practical experiments. For the time-of-flight and
noise-correlation detections, in the practical experiments where
the shallow harmonic trapping potential exists, we may obtain
lower peaks at the edge points than at the center in the expanding
atomic cloud.

 In summary, we have investigated the quantum phases of
ultracold bosonic atoms  in a two-dimensional optical
superlattice. First, we  derived the extended Bose-Hubbard
Hamiltonian without nearest-neighbor interactions. In our model,
through solving the extended Bose-Hubbard Hamiltonian with
Gutzwiller approach we found that the modulated superfluid,
density wave, and  Mott-insulator phases  appear. The detection of
these quantum phases with the time-of-flight and noise-correlation
techniques was proposed. We have calculated  the first order and
second order correlation functions for different quantum phases
and discussed their differences to distinguish them in
experiments.

\begin{acknowledgments} This work was  supported  by the
Teaching and Research Foundation for the Outstanding Young Faculty
of Southeast University  and  NSF of China Grant No. 10571091.
\end{acknowledgments}

\newpage

\begin{thebibliography}{99}
\bibitem{Jaksch} D. Jaksch, C. Bruder, J. I. Cirac, C. W. Gardiner, P. Zoller, Phys. Rev. Lett. 81 (1998) 3108.
\bibitem{Greiner}M. Greiner, O. Mandel, T. Esslinger, T. W. H\"ansch and I. Bloch, Nature 415 (2002) 39.
\bibitem{Paredes}B. Paredes, A. Widera, V. Murg, O. Mandel, S. F\"olling, I. Cirac, G. V. Shlyapnikov, T. W. H\"ansch,  I. Bloch, Nature 429 (2004) 277.
\bibitem{Stoferle}T. St\"oferle, H. Moritz, C. Schori, M. K\"ohl, T. Esslinger,  Phys. Rev. Lett. 92 (2004) 130403.
\bibitem{Fallani}L. Fallani,  J. E. Lye, V. Guarrera, C. Fort, M. Inguscio, Phys. Rev. Lett. 98 (2007) 130404.
\bibitem{Winkler}K. Winkler, G. Thalhammer, F. Lang, R. Grimm, J. Hecker Denschlag, A. J. Daley, A. Kantian, H. P. B\"uchler, P. Zoller, Nature 441 (2006) 853.
\bibitem{Gunter} K. G\"unter, T. St\"oferle, H. Moritz, M. K\"ohl, T. Esslinger, Phys. Rev. Lett. 96 (2006)
180402.
\bibitem{Ospelkaus}S. Ospelkaus, C. Ospelkaus, O. Wille, M. Succo, P. Ernst, K. Sengstock, .K. Bongs, Phys. Rev. Lett. 96  (2006) 180403.
\bibitem{Zwierlein}M.W. Zwierlein, C.n H. Schunck, A. Schirotzek, W. Ketterle, Nature 442 (2006) 54.
\bibitem{Chin}J. K. Chin, D. E. Miller, Y. Liu, C. Stan, W. Setiawan, C. Sanner, K. Xu, W. Ketterle,  Nature 443 (2006) 961.
\bibitem{Kohl} M.
K\"ohl, H. Moritz, T. St\"oferle, K. G\"unter, T. Esslinger, Phys.
Rev. Lett. 94 (2005) 080403.
\bibitem{Stoferle2}T. St\"oferle, H. Moritz, K. G\"unter, M. K\"ohl, T. Esslinger, Phys. Rev. Lett. 96 (2006) 030401.
\bibitem{Rom} T. Rom, T. Best, D. van Oosten, U. Schneider, S. F\"olling, B. Paredes, I. Bloch, Nature 444 (2006) 733.
\bibitem{He}P. B. He, Q. Sun, P. Li, S. Q. Shen, W. M. Liu, Phys. Rev. A 76, 043618
(2007).
\bibitem{Xie}Z. W. Xie, Z. X. Cao, E. I. Kats, W. M. Liu, Phys. Rev. A 71, 025601
(2005).
\bibitem{Zheng}G. P. Zheng, J. Q. Liang, W. M. Liu, Phys. Rev. A 71, 053608
(2005).
\bibitem{Li}Z. D. Li, P. B. He, L. Li, J. Q. Liang, W. M. Liu, Phys. Rev. A 71,
053611(2005).
\bibitem{Jaksch1}D. Jaksch, H.-J. Briegel, J. I. Cirac, C. W. Gardiner, P. Zoller, Phys. Rev. Lett. 82 (1999)
1975.
\bibitem{Brennen}G. K. Brennen, C. M. Caves, P. S. Jessen, I. H. Deutsch, Phys. Rev. Lett. 82 (1999)
1060.
\bibitem{Pachos} J. K. Pachos and P. L. Knight, Phys. Rev. Lett.
91 (2003) 107902.

 \bibitem{Buonsante}Buonsante, A. Vezzani, Phys. Rev. A 70 (2004)
 033608.
 \bibitem{Buonsante2}
 Buonsante, V. Penna, A. Vezzani, Phys. Rev. A 70 (2004) 061603.
 \bibitem{Buonsante3} Buonsante V. Penna, A. Vezzani, Laser Phys. 15 (2005)
 361.
 \bibitem{Buonsante4}Buonsante, A. Vezzani, Phys. Rev. A 72 (2005)
 013614.
 \bibitem{Buonsante5}
Buonsante,  V. Penna, A. Vezzani, Phys. Rev. A 72 (2005) 031602.
\bibitem{Rousseau} V. G. Rousseau, D. P. Arovas, M. Rigol, F. H\'ebert, G. G. Batrouni, R. T. Scalettar, Phys. Rev. B 73 (2006)
174516.
\bibitem{Altman} E. Altman, E. Demler, M. D. Lukin,
Phys. Rev. A 70 (2004) 013603.
\bibitem{Scarola}V. W. Scarola, E. Demler, S. Das Sarma, Phys. Rev. A 73 (2006) 051601(R).
\bibitem{Greiner1} M. Greiner, C. A. Regal, J. T. Stewart, D. S. Jin, Phys. Rev. Lett. 94 (2005) 110401.

\end{thebibliography}
\end{document}